\documentclass[11pt]{article}
\usepackage{amsfonts}
\usepackage{graphics}
\usepackage{psfrag}
\usepackage{epsfig}
\usepackage{epsf}
\usepackage{float}
\usepackage{amssymb,stmaryrd,latexsym}
\usepackage{amsmath}
\newcommand{\be}{\begin{equation}}
\newcommand{\ee}{\end{equation}}
\newcommand{\ba}{\begin{eqnarray}}
\newcommand{\ea}{\end{eqnarray}}
\newcommand{\non}{\nonumber}
\newcommand{\cirm}{\overset{_\circ}{m}}

\begin{document}

\title{\vspace{-1.2cm}
\hfill {\tiny FZJ--IKP(TH)--2008--14, HISKP-TH-08-15}\\
\vspace{1cm}%
Mass splittings within heavy baryon isospin multiplets\\ in 
chiral perturbation theory}

\author{Feng-Kun Guo$^{\rm a}$\footnote{{\it Email address:} f.k.guo@fz-juelich.de},
Christoph Hanhart$^{\rm a}$\footnote{{\it Email address:}
c.hanhart@fz-juelich.de},
and Ulf-G. Mei{\ss}ner$^{\rm a, b}$\footnote{{\it Email address:} meissner@itkp.uni-bonn.de}%
\\[2mm]
{\small $^{\rm a}${\it Institut f\"{u}r Kernphysik and J\"ulich Center for
    Hadron Physics,}}\\
{\small {\it Forschungszentrum J\"{u}lich, D--52425 J\"{u}lich, Germany}} \\[0mm]
{\small $^{\rm b}${\it Helmholtz-Institut f\"ur Strahlen- und
Kernphysik
(Theorie) and Bethe Center for Theoretical Physics,}}\\
{\small {\it Universit\"at Bonn, D--53115 Bonn, Germany}} }

\date{}

\maketitle

\begin{abstract}
\noindent
  We calculate the mass splittings within the heavy baryon isospin
  multiplets $\Sigma_{c(b)}$ and $\Xi_{c(b)}'$ in chiral
  perturbation theory to leading one--loop order. The pattern of the
  mass splittings in the $\Sigma_c$ iso-triplet, which is different
  from that of any other known isospin multiplet, can be explained. We
  predict $m_{\Xi_c'^+} - m_{\Xi_c'^0}=-0.2\pm0.6$~MeV, $m_{\Xi_b'^0}
  - m_{\Xi_b'^-}=-4.0\pm1.9$~MeV and the mass of the $\Sigma_b^0$ to
  be $5810.3\pm1.9~{\rm MeV}$.
\end{abstract}



\vspace{1cm}

\section{Introduction}

Mass splittings within isospin multiplets of hadrons appear due to
both the mass difference between the $u$ and $d$ quarks and
electromagnetic (em) effects. Since the $d$ quark is heavier than
the $u$, usually the hadron with more $d$ quarks is heavier within
one isospin multiplet. For instance, the neutron ($udd$) is heavier
than the proton ($uud$), and the $K^0(d{\bar s})$ is heavier than
the $K^+(u{\bar s})$. There is only one exception to this pattern,
the $\Sigma_c$ iso-triplet, consisting of the $\Sigma_c^{++}(cuu)$,
the $\Sigma_c^{+}(cud)$ and the $\Sigma_c^{0}(cdd)$. The mass splittings
within the $\Sigma_c$ iso-triplet are measured~\cite{Amsler:2008zz}
\begin{eqnarray}
\label{eq:deltamc} \Delta_{1c} &\!\!\!\equiv&\!\!\!
m_{\Sigma_c^{+}}-m_{\Sigma_c^0} = -0.9\pm0.4~{\rm MeV}, \non\\
\Delta_{2c} &\!\!\!\equiv&\!\!\! m_{\Sigma_c^{++}}-m_{\Sigma_c^0} =
0.27\pm0.11~{\rm MeV}\, .
\end{eqnarray}
Remarkably, the state with two $u$ quarks has the largest and the one with a $u$
and a $d$ quark has the smallest mass.

Only recently  some of the bottom cousins of the $\Sigma_c$, the
$\Sigma_b^{\pm}$, were observed by the CDF
Collaboration~\cite{:2007rw}. Their masses are for the $buu$ state
$m_{\Sigma_b^+}=5807.8\pm2.7$~MeV and for the $bdd$ state
$m_{\Sigma_b^-}=5815.2\pm2.0$~MeV, respectively ---  their neutral
partner $\Sigma_b^0$ has not been observed yet. Thus, here the
natural ordering of the states seems to be restored. On the other
hand, heavy quark symmetry relates baryons containing a $b$ quark to
those with a $c$ quark, which makes this different pattern even more
puzzling.  In this work we investigate the origin of
these patterns, together with those in the $\Xi_{c(b)}'$ doublets,
using chiral perturbation theory (CHPT) and heavy quark symmetry. In
this way we can include both sources of isospin violation in a way
consistent with QCD~\cite{Gasser:1983yg}. Our work is a
straightforward extension of analogous studies for the
nucleon~\cite{Meissner:1997ii,Muller:1999ww} (for pioneering
studies, see~\cite{Weinberg:1977hb,Gasser:1980sb}).

The isospin splittings for heavy baryons were already studied in
various quark models in
Refs.~\cite{Hwang:2008dj,Tiwari:1985ru,Capstick:1987cw,%
Chan:1985ty,Hwang:1986ee,Verma:1988gg,Cutkosky:1993cc,Varga:1998wp,SilvestreBrac:2003kd}.
The advantages of our investigation are that $(i)$, since we use an
effective field theory, the theoretical uncertainty can be
estimated, $(ii)$ for the first time meson loop corrections are
considered --- they turn out to be numerically significant for the
charm baryons, and $(iii)$ this study investigates all isospin
splittings at the same time. We compare our results to those of the
quark models below.

Our paper is organized as follows. In Section~\ref{sec:lag}, we
construct the SU(3) chiral effective Lagrangians responsible for the
mass corrections of the sextet heavy baryons at order
${\cal{O}}(p^2)$. Here, $p$ denotes the expansion parameter of the
underlying effective field theory.
The strong Lagrangian is proportional to the quark
mass and the em Lagrangian is constructed including virtual photons.
As we will show, the operator structure of the effective field
theory for the heavy--light baryons is richer than the one of the
light quark sector since heavy and light quarks must be treated
differently. Calculations up to ${\cal{O}}(p^3)$ are performed in
Section~\ref{sec:loop}. A brief summary is given in the last
section. Some technicalities are relegated to the appendices.

\section{The chiral effective Lagrangians}
\label{sec:lag}

In CHPT the quark mass difference enters explicitly through the
quark mass matrix. The inclusion of the virtual photons has been
first considered systematically for the three-flavor case in
Ref.~\cite{Urech:1994hd}. Chiral Lagrangians with virtual photons
have been constructed for the study of isospin symmetry breaking
phenomena in mesons and baryons with light up and down quarks~(see,
e.g.,
\cite{Meissner:1997ii,Muller:1999ww,Meissner:1997fa,Knecht:1997jw,
Fettes:2001cr,Kubis:1999db,Amoros:2001cp,Kubis:2001ij,Nehme:2001wf,Knecht:2002gz,
Gasser:2002am,Schweizer:2002ft,Gasser:2003hk,Bijnens:2007xa} for an
incomplete list). Recently, this technique was used to study the
interaction between Goldstone bosons and heavy--light mesons and the
isospin breaking decay width of the
$D_{s0}^*(2317)$~\cite{Guo:2008gp}.

In this paper, we use the technique of SU(2) chiral perturbation
theory to study the mass splittings within the heavy baryon isospin
multiplets. The $u,d$ quark masses and the electric charge $e$ are
counted as small quantities. They are booked as $m_{u},m_{d} \sim
{\cal O}(p^2)$, and $e\sim {\cal O}(p)$ as usual, where $p$ denotes
a small momentum with respect to the typical hadronic scale of about
1~GeV. Both types of isospin--violating effects $\sim (m_u-m_d)$ and
$\sim e$ are taken into account here in a systematic manner to the
order ${\cal{O}}(p^3)$. These effects can be accounted for in
three--flavor CHPT to study the mass splittings within the SU(3)
multiplets systematically. However, similar to the case of CHPT for
light baryons, the SU(3) breaking contributions from the
kaon--baryon and eta--baryon loops are
large, see e.g.~\cite{Bijnens:1985kj,Borasoy:1996bx,Donoghue:1998bs}, which
makes the convergence of the chiral expansion problematic. We will
therefore treat each heavy baryon isospin multiplet separately to
${\cal{O}}(p^3)$ in two-flavor CHPT, and relate the low--energy
constants (LECs) to ${\cal{O}}(p)$ and ${\cal{O}}(p^2)$ through
SU(3) relations (see also the discussion in Section~\ref{sec:loop}).
On the other hand, the splittings within the multiplets are well
behaved. This procedure is equivalent to starting from the SU(3) Lagrangian
and to calculate only the leading order SU(2) loops, i.e., the pion--baryon
loops.

In this section, we will construct the SU(3) chiral effective
Lagrangians pertinent to the mass corrections of the heavy baryons
to order ${\cal{O}}(p^3)$  (for similar works considering the
strange quark as heavy, see
Refs.~\cite{Roessl:1999iu,Frink:2002ht,Beane:2003yx,Tiburzi:2008bk}).

In order to construct the chiral effective Lagrangians, the
following building blocks are necessary (we employ the standard
nonlinear realization of  chiral symmetry)
\ba%
U &\!\!\!=&\!\!\! \exp \left( {\sqrt{2}i\phi \over
F_\pi}\right),\quad u^2=U,
\non\\
u_{\mu} &\!\!\!=&\!\!\! iu^{\dag}\nabla_{\mu}Uu^{\dag}, \non\\
\nabla_{\mu}U &\!\!\!=&\!\!\! \partial_\mu U-iQA_\mu U+iUQA_\mu,
\non\\
\chi_+ &\!\!\!=&\!\!\! u^\dagger \chi u^\dagger + u\chi u ,\non\\
Q_\pm &\!\!\!=&\!\!\! \frac12\left( u^\dagger Q u \pm uQu^\dagger
\right),
\ea%
where $F_{\pi}$ is the pion decay constant\footnote{Strictly speaking,
  this should be the pion decay constant in the chiral limit. To the
  accuracy we are working, however, we do not need to differentiate
  this from its physical value.}, $\phi$ collects the
Goldstone boson fields
\begin{eqnarray}
\label{eq:phi}
 \phi =
  \left(
    \begin{array}{c c c}
 \frac{1}{\sqrt{2}}\pi^0 + \frac{1}{\sqrt{6}}\eta & \pi^+ & K^+\\
\pi^- & - \frac{1}{\sqrt{2}}\pi^0 + \frac{1}{\sqrt{6}}\eta & K^0 \\
K^- & {\bar K}^0 & - \frac{2}{\sqrt{6}}\eta
    \end{array}
\right) ,
\end{eqnarray}
and $A_\mu$ is the photon (em) field.
The diagonal quark mass matrix and the charge matrix are
\ba%
\chi &\!\!\!=&\!\!\! 2B_0\cdot {\rm diag}\left\{m_u,m_d,
m_s\right\}, \non\\
Q &\!\!\!=&\!\!\! e \cdot{\rm diag}\left\{2/3,-1/3, -1/3\right\},
\ea%
in terms of $B_0= |\langle 0 |\bar q q |0\rangle|/F_{\pi}^2 $ and
the elementary electric charge $e$ ($e>0$). Under
SU(3)$_L\times$SU(3)$_R$, $u_\mu,\chi_+$ and $Q_{\pm}$ transform as
\be%
{\cal O} \to h {\cal O} h^{\dag},
\ee%
where the compensator field $h$ is an element of the conserved
vector subgroup $SU(3)_V$.

The $\Sigma_c$ iso-triplet and the $\Xi_{c}'$ iso-doublet belong to
the symmetric sextet and the $\Lambda_c^+$ and the $\Xi_{c}$ doublet
belong to the anti-symmetric triplet in the flavor SU(3)
classification. We use the following matrix representation in
accordance with the notation of Refs.~\cite{Yan:1992gz,Cheng:1993kp}
\begin{eqnarray}
B_{6c} = \frac1{\sqrt{2}}\left(
\begin{array}{c c c}
\sqrt{2}\Sigma_c^{++} & \Sigma_c^{+} & \Xi_c'^{+} \\
\Sigma_c^{+} & \sqrt{2}\Sigma_c^{0} & \Xi_c'^{0} \\
\Xi_c'^{+} & \Xi_c'^{0} & \sqrt{2}\Omega_c^0 \\
\end{array}
\right) , \quad %
B_{{\bar 3}c} = \left(
\begin{array}{c c c}
0 & \Lambda_c^+ & \Xi_c^{+} \\
-\Lambda_c^+ & 0 & \Xi_c^{0} \\
-\Xi_c^{+} & -\Xi_c^{0} & 0 \\
\end{array}
\right).
\end{eqnarray}
Under SU(3)$_L\times$SU(3)$_R$, the transformation laws of the
charmed baryon fields are~\cite{Yan:1992gz}
\begin{equation}
B_{6c}\to hB_{6c} h^T,\quad B_{{\bar 3}c}\to hB_{{\bar 3}c} h^T,
\end{equation}
where $h^T$ is the transpose of $h$. The matrices
for the bottom baryons can be obtained replacing $c$ by $b$ and
decreasing the electric charge of every state by one unit.

Let $B_{{\bar 3}Q},~B_{6Q}$~($Q=c,b$) denote the heavy baryon
fields, and $\cirm_{{\bar 3}Q},~\cirm_{6Q}$ their masses in the
chiral limit, respectively. Analogous to the effective chiral
Lagrangian for the pion--nucleon system, the lowest order
Lagrangian involving the sextet and anti-triplet heavy baryon fields
can be written as~\cite{Yan:1992gz,Cho:1992gg}
\begin{eqnarray}
\label{eq:L1} {\cal L}^{(1)} &\!\!\!=&\!\!\! {1\over 2}\left\langle
{\bar B_{{\bar 3}Q}} (i\gamma_{\mu}\tilde{D}^{\mu}-\cirm_{{\bar
3}_Q}) B_{{\bar 3}Q}\right\rangle + \left\langle {\bar B_{6Q}}
(i\gamma_{\mu}\tilde{D}^{\mu}-\cirm_{6_Q})
B_{6Q}\right\rangle \non\\
&\!\!&\!\! + {1\over 2}g_1 \left\langle {\bar B_{6Q}}
\gamma_{\mu}\gamma_5\tilde{u}^{\mu} B_{6Q}\right\rangle + {1\over
2}g_2 \left\langle {\bar B_{6Q}} \gamma_{\mu}\gamma_5\tilde{u}^{\mu}
B_{{\bar 3}Q}\right\rangle + h.c. + \frac12g_3 \left\langle {\bar
B_{{\bar3}Q}} \gamma_{\mu}\gamma_5\tilde{u}^{\mu} B_{{\bar
3}Q}\right\rangle.
\end{eqnarray}
with
\begin{equation}
\tilde{D}_\mu = D_\mu - iQ_{B+}A_\mu, \quad \tilde{u}_\mu = u_\mu -
2Q_{B-}A_\mu~.
\end{equation}
The chiral covariant derivative on the baryon fields $D_\mu$ is given by
\begin{eqnarray}
D_{\mu} &\!\!=&\!\! \partial_{\mu}+\Gamma_{\mu}, \nonumber\\
\Gamma_{\mu} &\!\!=&\!\!
{1\over2}\left(u^{\dag}\partial_{\mu}u+u\partial_{\mu}u^{\dag}\right).
\end{eqnarray}
The charge matrix of the heavy baryons $Q_B$, which gives the
correct minimal coupling of the heavy baryons to photons, is
constructed as
\be%
Q_B = 2Q + q_h\mathbb{I}=
\begin{cases}
e\cdot {\rm diag}\left\{2,0,0\right\},
& \text{for the charm baryons,} \\
 e\cdot {\rm diag}\left\{1,-1,-1\right\}, &
\text{for the bottom baryons,}
\end{cases}
\ee%
where $q_h$ is the charge of the heavy quark, and $\mathbb{I}$ is a
$3\times3$ unit matrix. $Q_{B\pm}$ is defined as
$$Q_{B\pm} = \frac12\left(u^\dagger Q_B u \pm uQ_Bu^\dagger\right).$$

At ${\cal{O}}(p^2)$ the strong Lagrangian pertinent to the
corrections of the masses is given by the terms containing one power
of the quark mass matrix
\begin{equation}
\label{eq:L2str} {\cal L}^{(2)}_{\rm str.} = - \left\langle {\bar
B_Q} \left(\alpha_1\chi_+ + \alpha_2 \left\langle \chi_+
\right\rangle \right) B_Q\right\rangle.
\end{equation}
The em Lagrangian at  ${\cal{O}}(p^2)$, parameterizing hard virtual
photons, is more complicated than the one of the pion--nucleon
system~\cite{Meissner:1997ii,Muller:1999ww}. The terms, which
contribute to the mass corrections of the sextet heavy baryons and
are quadratic in the light quark charge matrix read
\begin{eqnarray}
\label{eq:L2em} {\cal L}^{(2)}_{QQ} &\!\!\!=&\!\!\! - F_{\pi}^2
\left\langle {\bar B_{6Q}} \left[ \beta_0\left(Q_+^2-Q_-^2\right) +
\beta_1Q_+\left\langle Q_+ \right\rangle + \beta_2\left\langle
Q_+^2-Q_-^2 \right\rangle + \beta_3\left\langle Q_+^2+Q_-^2
\right\rangle \right]
B_{6Q}\right\rangle \non\\
&\!\!\!&\!\!\! -F_\pi^2\beta_4 \left\langle Q_+^T {\bar B_{6Q}}Q_+
B_{6Q} \right\rangle.
\end{eqnarray}
The em Lagrangian given above only deals
with the hard photons exchanged between the light quarks in the heavy
baryons. In addition we need to add terms that parameterize
the em interactions between the heavy and a light quark.
Since the heavy quark can be viewed as static, its charge $q_h$
acts as a static background field that transforms as a
scalar under
SU(3)$_L\times$SU(3)$_R$.
It is easy to see that in this way we get one
additional independent structure that contributes to
the isospin splittings, namely
\be%
{\cal L}^{(2)}_{\rm em} = {\cal L}^{(2)}_{QQ} - F_{\pi}^2\beta_{1h}
\left\langle {\bar B_{6Q}}Q_+\langle q_h\mathbb{I}\rangle B_{6Q}
\right\rangle.
\ee%
As we will show below, it is this term which makes the mass
splitting pattern within the $\Sigma_c$ iso-triplet different from
that within the $\Sigma_b$ iso-triplet. The analogous mechanism is
also present in the mentioned quark model calculations.

\section{Mass splittings within the heavy baryon isospin multiplets}
\label{sec:loop}

As mentioned above, our strategy is to
relate the LECs for different heavy baryon isospin multiplets
using SU(3) relations,
but, since we are only after isospin splittings,
 calculate only the pion--baryon loop
corrections. If we start from the SU(2) Lagrangians for each of the
isospin multiplets, the chiral limit masses of the $\Sigma_Q$ and
$\Xi_Q'$ are different, and the difference contributes to the SU(3)
mass splittings of ${\cal{O}}(p^3)$. It is important to stress that
the differences of the LECs between their SU(2) and SU(3) values
starts contributing at order ${\cal{O}}(p^4)$ and thus is beyond the
accuracy of our calculation.\footnote{The precise matching between
two-- and three--flavor versions of CHPT is discussed in
Refs.~\cite{Gasser:1984gg,Frink:2004ic,Gasser:2007sg}.} Therefore,
what we do is equivalent to use the SU(3) Lagrangians given in
Section~\ref{sec:lag} to do SU(2) calculations. Indeed, the number
of independent operators matches for the two cases, since the
contribution of the $\beta_1$ term to the baryon masses vanishes for
SU(3) because the three--flavor quark charge matrix is traceless,
and the $\beta_0$ term for SU(3) can be translated into the
$\beta_1$ term for SU(2) by using the Cayley--Hamilton relation for
$2\times2$ matrices
\begin{equation}
Q_+^2-Q_-^2 = Q_+\left\langle Q_+ \right\rangle + {1\over 2} \left(
\left\langle Q_+^2-Q_-^2 \right\rangle - \left\langle Q_+
\right\rangle^2 \right)~.
\end{equation}

\subsection{Mass splittings to ${\cal{O}}(p^2)$}

At leading order, there is no mass splitting within isospin
multiplets. However, at next--to--leading order the terms in the
${\cal{O}}(p^2)$ Lagrangians generate mass splittings. From
Eqs.~(\ref{eq:L2str},\ref{eq:L2em}), we get
\begin{eqnarray}
\label{eq:delc2} \Delta_{1c}^{(2)} &\!\!\!\equiv&\!\!\!
\left(m_{\Sigma_c^+}-m_{\Sigma_c^0}\right)^{(2)} = 2
\alpha_1B_0(m_u-m_d)  + {1\over 6} F_{\pi}^2e^2 (\beta_0-2\beta_4+6\beta_{1h}), \non\\
\Delta_{2c}^{(2)} &\!\!\!\equiv&\!\!\!
\left(m_{\Sigma_c^{++}}-m_{\Sigma_c^0}\right)^{(2)} = 4
\alpha_1B_0(m_u-m_d) + {1\over 3} F_{\pi}^2e^2
(\beta_0+\beta_4+6\beta_{1h}), \non\\
\Delta_{3c}^{(2)} &\!\!\!\equiv&\!\!\!
\left(m_{\Xi_c'^{+}}-m_{\Xi_c'^0}\right)^{(2)} = 2
\alpha_1B_0(m_u-m_d) + {1\over 6} F_{\pi}^2e^2
(\beta_0-2\beta_4+6\beta_{1h})  = \Delta_{1c}^{(2)},
\end{eqnarray}
for the charm baryons $\Sigma_c$ and $\Xi_c'$. Similarly, for the
bottom baryons we have
\begin{eqnarray}
\label{eq:delb2} \Delta_{1b}^{(2)} &\!\!\!\equiv&\!\!\!
\left(m_{\Sigma_b^0}-m_{\Sigma_b^-}\right)^{(2)} = 2
\alpha_1B_0(m_u-m_d)  + {1\over 6} F_{\pi}^2e^2 (\beta_0-2\beta_4-3\beta_{1h}), \non\\
\Delta_{2b}^{(2)} &\!\!\!\equiv&\!\!\!
\left(m_{\Sigma_b^{+}}-m_{\Sigma_b^-}\right)^{(2)} = 4
\alpha_1B_0(m_u-m_d) + {1\over 3} F_{\pi}^2e^2
(\beta_0+\beta_4-3\beta_{1h}), \non\\
\Delta_{3b}^{(2)} &\!\!\!\equiv&\!\!\!
\left(m_{\Xi_b'^{0}}-m_{\Xi_b'^-}\right)^{(2)} = 2
\alpha_1B_0(m_u-m_d) + {1\over 6} F_{\pi}^2e^2
(\beta_0-2\beta_4-3\beta_{1h})  = \Delta_{1b}^{(2)}.
\end{eqnarray}
Note that the $\alpha_1$ and $\beta_0$ terms always appear in the
same linear combination, which means that the strong contribution,
the $\alpha_1$ term, cannot be disentangled from the em contribution
without additional information. This is completely analogous to the
case of the neutron--proton mass splitting, see e.g.
Ref.~\cite{Meissner:1997ii}. The $\beta_{1h}$ term has a different
sign for charm baryons and bottom baryons, hence it is expected to
induce a different interference pattern of the various
contributions.

From the relations given above we find
\ba%
\label{eq:delta0} \left(m_{\Xi_c'^{+}}-m_{\Xi_c'^{0}}\right) -
\left(m_{\Sigma_c^{+}} - m_{\Sigma_c^{0}}\right) &\!\!\!=&\!\!\! {\cal O}(p^3), \non\\
\left(m_{\Xi_b'^{0}}-m_{\Xi_b'^{-}}\right) - \left(m_{\Sigma_b^{0}}
- m_{\Sigma_b^{-}}\right)
&\!\!\!=&\!\!\! {\cal O}(p^3), \non\\
\left(m_{\Sigma_b^{+}} + m_{\Sigma_b^{-}} - 2m_{\Sigma_b^{0}}\right)
- \left(m_{\Sigma_c^{++}} + m_{\Sigma_c^{0}} -
2m_{\Sigma_c^{+}}\right) &\!\!\!=&\!\!\! {\cal O}(p^3).
\ea%
The last relation is obtained invoking heavy quark symmetry.

\subsection{Mass splittings to ${\cal{O}}(p^3)$}

The first non-vanishing loop corrections to the baryon masses appear
at order ${\cal{O}}(p^3)$. At this order formally both photon loops
as well as pion--baryon loops contribute. Remarkably, QCD does not
allow for a counterterm at this order and consequently the
${\cal{O}}(p^3)$ pieces of the these loops are finite. We start with
the latter kind of loops that are to be constructed from  two
vertices of ${\cal{O}}(p)$. A complete list of loops is given in
Table~\ref{tab:loops} for the charm baryons considered here.
\begin{table}[t]
\begin{center}
\begin{tabular}{|l l|}\hline\hline
Baryons & Loops \\ \hline\\[-4mm]%
$\Sigma_c^{++}$ & $\Sigma_c^{++}\pi^0$, $\Sigma_c^{+}\pi^+$, $\Lambda_c^{+}\pi^+$ \\
$\Sigma_c^{+}$ & $\Sigma_c^{++}\pi^-$, $\Sigma_c^{0}\pi^+$, $\Lambda_c^{+}\pi^0$ \\
$\Sigma_c^{0}$ & $\Sigma_c^{+}\pi^-$, $\Sigma_c^{0}\pi^0$, $\Lambda_c^{+}\pi^-$ \\
$\Xi_c'^{+}$ & $\Xi_c'^{+}\pi^0$, $\Xi_c'^{0}\pi^+$, $\Xi_c^{+}\pi^0$, $\Xi_c^{0}\pi^+$ \\
$\Xi_c'^{0}$ & $\Xi_c'^{+}\pi^-$, $\Xi_c'^{0}\pi^0$, $\Xi_c^{+}\pi^-$, $\Xi_c^{0}\pi^0$ \\ \hline\hline%
\end{tabular}
\caption{\label{tab:loops}Pion--baryon loops contributing to the
charm baryon mass corrections.}
\end{center}
\end{table}
\begin{figure}[htb]
\begin{center}
\epsfig{file=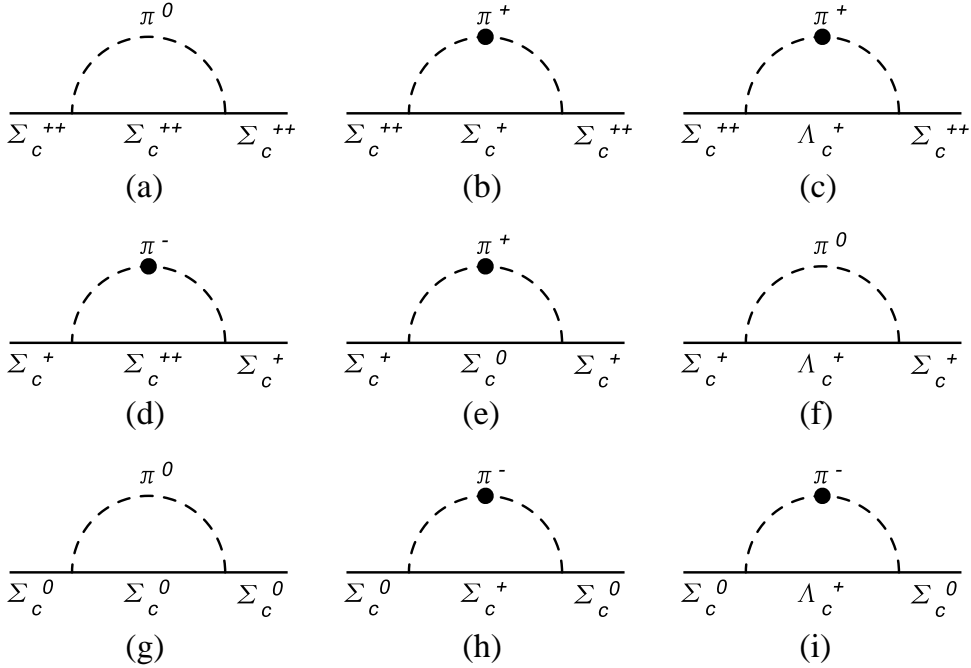, width=0.8\textwidth} \caption{The
pion--baryon loops contributing to the self-energies of the
$\Sigma_c^{++}$ (a, b, c), $\Sigma_c^{+}$ (d, e, f) and
$\Sigma_c^{0}$ (g, h, i). The black dots in the charged pion
propagators denote the electromagnetic insertions at
${\cal{O}}(p^2)$. \label{fig:pionloop}}
\end{center}
\end{figure}
Since in the power counting the pion mass difference is of the same
order as the pion mass itself, in the loops we are to use the
physical pion masses (for a detailed discussion of this point, see
e.g. Ref.~\cite{Gasser:2002am}). On the other hand, to the order we
are working, the masses to be used for the baryons of the same
isospin multiplet are the same. Therefore, the leading SU(2) loop
contributions to the mass corrections of the $\Xi_c'^{+}$ and
$\Xi_c'^{0}$ as well as $\Sigma_c^{++}$ and $\Sigma_c^{0}$ are
equal, as can be seen from Table~\ref{tab:loops}. Hence there is no
loop correction for the corresponding mass differences at
${\cal{O}}(p^3)$. The pion--baryon loops for the $\Sigma_c$
self-energies are shown in Fig.~\ref{fig:pionloop}. Contrary to the
case of the nucleon mass differences, here it is not straightforward
to use the heavy baryon formalism to calculate the pion--baryon
loops, since the pion--$\Lambda_c$ contribution generates a cut. It
is thus more convenient for us to evaluate the integrals using the
covariant method of infrared regularization as derived by Becher and
Leutwyler~\cite{Becher:1999he}.  Some remarks on the method and the
relevant integrals are given in Appendix~\ref{app:ir}.

\begin{figure}[t]
\begin{center}
\epsfig{file=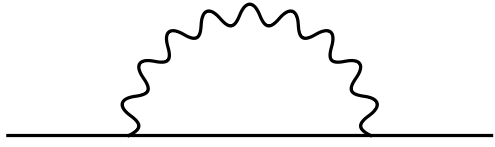, width=0.4\textwidth} \caption{The
photon--baryon loop. \label{fig:photonloop}}
\end{center}
\end{figure}

The photon--baryon loops are shown in  Fig.~\ref{fig:photonloop}.
Formally they also contribute at ${\cal{O}}(p^3)$. However, it can
be shown that they vanish to this order. Since the baryon mass in
the loop is equal to the mass of the external legs, for this loop we
may use the integral representation of the heavy  baryon formalism.
Then the vanishing of the loop follows from the absence of a mass
scale in the integral. This result also holds, when the infrared
regularization is employed, as outlined in Appendix~\ref{app:ir}.

Therefore the mass splittings for the charm baryons $\Sigma_c$ and $\Xi_c'$
are to third order in the chiral expansion
\ba%
\label{eq:rescharm} m_{\Sigma_c^+}-m_{\Sigma_c^0}
&\!\!\!\equiv&\!\!\! \Delta_{1c}^{(2)}  +
\Delta_{1c}^{\rm loop}(m_{\Sigma_c},m_{\Lambda_c}) + {\cal O}(p^4) \ , \\
 m_{\Sigma_c^{++}}-m_{\Sigma_c^0} &\!\!\!\equiv&\!\!\! \Delta_{2c}^{(2)} + {\cal O}(p^4) \ , \\
 m_{\Xi_c'^{+}}-m_{\Xi_c'^0} &\!\!\!\equiv&\!\!\! \Delta_{1c}^{(2)} + {\cal O}(p^4) \ .
\ea%
The mass splittings for the bottom baryons $\Sigma_b$ and
$\Xi_b'$ are
\ba%
 m_{\Sigma_b^0}-m_{\Sigma_b^-} &\!\!\!\equiv&\!\!\! \Delta_{1b}^{(2)} + \Delta_{1b}^{\rm loop}(m_{\Sigma_b},m_{\Lambda_b})
 + {\cal O}(p^4) \ , \\
 m_{\Sigma_b^{+}}-m_{\Sigma_b^-} &\!\!\!\equiv&\!\!\! \Delta_{2b}^{(2)} + {\cal O}(p^4) \ , \\
 m_{\Xi_b'^{0}}-m_{\Xi_b'^-} &\!\!\!\equiv&\!\!\! \Delta_{1b}^{(2)} + {\cal O}(p^4) \ ,
 \label{eq:resbottom}
\ea%
where the results to ${\cal O}(p^2)$, the $\Delta^{(2)}$ were given
in the previous subsection. As derived in Appendix~\ref{app:loop},
the loop functions to ${\cal O}(p^3)$ are given by
\ba%
\non \Delta_{1Q}^{\rm
loop}(m_{\Sigma_Q},m_{\Lambda_Q})&\!\!\!{=}&\!\!\! -
\frac{g_1^2}{32\pi
  F_{\pi}^2}\left(M_{\pm}^3-M_0^3\right)\\
  &\!\!\! &\!\!\! + {\rm Re}\Sigma^{\rm
  (c)}(m_{\Sigma_Q};M_{0},m_{\Lambda_Q}) -{\rm Re}\Sigma^{\rm
  (c)}(m_{\Sigma_Q};M_{\pm},m_{\Lambda_Q}) \ ,
\ea%
with $M_0 \, (M_\pm)$ the neutral (charged) pion mass. The explicit
expressions for the loop functions are given in
Appendix~\ref{app:loop}.

\subsection{Numerical results}

When calculating the loops, we take the physical values for  the masses
\ba%
M_{\pm} &\!\!\!=&\!\!\! 139.57~{\rm MeV},\quad\, M_0 = 134.98~{\rm MeV}, \non\\
m_{\Sigma_c} &\!\!\!=&\!\!\! 2453.56~{\rm MeV},\quad m_{\Lambda_c} = 2286.46~{\rm MeV}, \non\\
m_{\Sigma_b} &\!\!\!=&\!\!\! 5811.5~{\rm MeV},\quad\,\, m_{\Lambda_b} =
5620.2~{\rm MeV}.
\ea%
The values of $g_1$ and $g_2$ can be estimated based on SU(6)~\cite{Yan:1992gz}
\begin{equation}
g_1 = {4\over3}g_A^{ud},\quad g_2 = -\sqrt{2\over3}g_A^{ud},
\end{equation}
where $g_A^{ud}$ is the coupling constant for the single quark
transition $u\to d$. A value of $g_A^{ud}=0.76$ gives the correct
nucleon axial coupling constant $g_A=(5/3)g_A^{ud}=1.27$,
correspondingly $g_1=1.02$. It also gives $g_2=-0.62$, the absolute
value of which is close to the empirical one $|g_2|=0.58\pm0.04$
obtained from the measured decay width
$\Gamma(\Sigma_c^{++}\to\Lambda_c^+\pi^+)=2.23\pm0.30$~MeV. The only
unknown parameters are the LECs in the ${\cal{O}}(p^2)$ Lagrangians. There
are effectively three
\ba%
\tilde{\gamma} &\!\!\!\equiv&\!\!\! 2 \alpha_1B_0(m_u-m_d) + \frac{1}{6}e^2F_\pi^2\beta_0,\non\\
\tilde{\beta}_4 &\!\!\!\equiv&\!\!\! e^2F_\pi^2\beta_4,\non\\
\tilde{\beta}_{1h} &\!\!\!\equiv&\!\!\!  e^2F_\pi^2\beta_{1h}.
\ea%
Totally there are four known isospin mass splittings of the sextet
heavy baryons,\footnote{The first two mass splittings are given in
PDG~\cite{Amsler:2008zz}. The last two mass splittings are evaluated
by taking the difference of the masses given in PDG:
$m_{\Sigma_b^+}=5807.8\pm2.7$~MeV,
$m_{\Sigma_b^-}=5815.2\pm2.0$~MeV, $m_{\Xi_c'^+}=2575.7\pm3.1$~MeV,
and $m_{\Xi_c'^0}=2578.0\pm2.9$~MeV.}
\ba%
m_{\Sigma_c^+} - m_{\Sigma_c^0} &\!\!\!=&\!\!\! - 0.9\pm0.4~{\rm MeV}, \non\\
m_{\Sigma_c^{++}} - m_{\Sigma_c^0} &\!\!\!=&\!\!\! 0.27\pm0.11~{\rm MeV}, \non\\
m_{\Sigma_b^+} - m_{\Sigma_b^-} &\!\!\!=&\!\!\! - 7.4\pm3.4~{\rm MeV}, \non\\
m_{\Xi_c'^+} - m_{\Xi_c'^0} &\!\!\!=&\!\!\! - 2.3\pm4.2~{\rm MeV}.
\ea%
The first three  will be taken to
determine the LECs $\tilde{\gamma},~\tilde{\beta}_4$ and
$\tilde{\beta}_{1h}$ because they have the smallest uncertainty.

Using the physical value of the pion decay constant
$F_\pi=92.4$~MeV, we get the contribution of the loops to
$m_{\Sigma_c^{+}}-m_{\Sigma_c^0}$
\ba%
\label{eq:g1g2} \Delta_{1c}^{\rm loop} &\!\!\!=&\!\!\!
(-0.32\pm0.15) + (-0.41\pm0.06)~{\rm MeV} \non\\
&\!\!\!=&\!\!\! -0.73\pm0.16~{\rm MeV},
\ea%
where in the first line the numbers in the first parenthesis are from
the $\pi$--$\Sigma_c$ loops, and those in the second parenthesis
from the $\pi$--$\Lambda_c$ loops.
The uncertainty of the loops is in general controlled by
the expansion parameter of CHPT, namely  $\chi=M/\Lambda_\chi$,
with the chiral symmetry breaking scale $\Lambda_\chi\simeq 1\,$GeV.
Since we do not really know the value of $g_1$, which enters
in the $\pi$--$\Sigma_c$ loops, for those we estimate the
uncertainty conservatively as being of the order $4\chi$.
Since the coupling constant of the $\Sigma_c$ to  $\Lambda_c$ and $\pi$
can be extracted from experiment, for the second contribution we use
directly the uncertainty that results from that extraction (see above).
Note that the given uncertainty is at the same time numerically of order $\chi$.
Thus the uncertainty estimate is consistent with what is expected from the chiral expansion.
The LECs are then determined as
\ba%
\label{eq:lec2num}
\tilde{\gamma} &\!\!\!=&\!\!\! -2.5\pm1.1~{\rm MeV},\non\\
\tilde{\beta}_4 &\!\!\!=&\!\!\! 0.6\pm0.9~{\rm MeV}, \non\\
\tilde{\beta}_{1h} &\!\!\!=&\!\!\! 2.6\pm1.1~{\rm MeV}.
\ea%
The two different contributions in Eq.~(\ref{eq:g1g2}) are
comparable, and they are considerably smaller than the individual
strong and em contributions at ${\cal{O}}(p^2)$, see Eqs.~(\ref{eq:delc2})
and (\ref{eq:lec2num}), showing good convergence of the chiral
expansion. The mass splitting within the $\Xi_c'$ doublet can be predicted
\be%
m_{\Xi_c'^+} - m_{\Xi_c'^0} = m_{\Sigma_c^+} - m_{\Sigma_c^0} -
\Delta_{1c}^{\rm loop} = -0.2\pm0.4(\rm exp)\pm0.4(\rm th)~{\rm
MeV},
\ee%
where the first uncertainty coming from the uncertainty of
$m_{\Sigma_c^+} - m_{\Sigma_c^0}$ is experimental, and the second
one is theoretical. It comes from neglecting the ${\cal{O}}(p^4)$
contribution and was estimated by taking
one half of the leading loop contribution.
\renewcommand{\arraystretch}{1.2}
\begin{table}[t]
\begin{center}
\begin{tabular}{|c|c|c|c|c|c|c|}\hline\hline
 & $\Sigma_c^+-\Sigma_c^0$ & $\Sigma_c^{++}-\Sigma_c^0$ & $\Xi_c'^+-\Xi_c'^0$
 & $\Sigma_b^0-\Sigma_b^-$ & $\Sigma_b^+-\Sigma_b^-$ & $\Xi_b'^0-\Xi_b'^-$ \\
\hline
Exp.~\cite{Amsler:2008zz} & $-0.9\pm0.4$ & $0.27\pm0.11$ & $-2.3\pm4.2$ &
& $-7.4\pm3.4$ &  \\
\hline
Our work & $-0.9\pm0.4^*$ & $0.27\pm0.11^*$ & $-0.2\pm0.6$ & $-4.9\pm1.9$ &
$-7.4\pm3.4^*$ & $-4.0\pm1.9$ \\
\hline
\cite{Hwang:2008dj} & $-0.9\pm0.4^*$ & $0.27\pm0.11^*$ &  & $-6.9\pm1.1$ &
$-7.4\pm2.3^*$ & \\
\cite{Tiwari:1985ru} & $-0.83$ & $-0.01$ & $-1.48$ &   &   &  \\%
\cite{Chan:1985ty} & $-0.73$ & 0.28 & $-3.20$ & $-3.95$ & $-6.12$ & $-6.16$ \\%
\cite{Capstick:1987cw} & $-0.2~$ & 1.4~ &  & $-3.7~$ & $-5.6~$ & \\%
\cite{Hwang:1986ee} & $-0.5~$ & 3.0~ & $-1.0~$ & $-5.6~$ & $-7.1~$ & \\%
\cite{Verma:1988gg} & $-0.7~$ & 0.5~ & $-1.2~$ &  &  & \\%
\cite{Cutkosky:1993cc} & $-0.40$ & 0.84 & & & & \\%
\cite{Varga:1998wp} & $-0.36$ & $1.20$ & $-0.30$ & $-2.51$ & $-3.57$ & \\%
\cite{SilvestreBrac:2003kd} & $-0.33$ & 0.37 & $-0.20$ & & & \\%
\hline\hline%
\end{tabular}
\caption{\label{tab:results}Comparison of our results with the
experimental data and the results from quark models (units are
MeV). The numbers marked by $*$ are used as inputs.}
\end{center}
\end{table}
Because the width of the $\Sigma_b$ has not been measured so far,
$g_2$ for the bottom baryons cannot be determined from the data.
Taking the same values as before for $g_1$ and $g_2$, as dictated by
heavy quark symmetry, we get the mass difference between
$\Sigma_b^0$ and $\Sigma_b^-$ at ${\cal{O}}(p^3)$
\ba%
\Delta_{1b}^{\rm loop} &\!\!\!=&\!\!\!
(-0.3\pm0.2) + (-0.6\pm0.1)~{\rm MeV} \non\\
&\!\!\!=&\!\!\! -0.9\pm0.2~{\rm MeV}.
\ea%
where in the first line the numbers in the first and the second
parentheses are from the $\pi$--$\Sigma_b$ loops and the
$\pi$--$\Lambda_b$ loops, respectively. The uncertainties were
estimated as in case of the charm baryons.
 According to
Eq.~(\ref{eq:delb2}), the mass of the $\Sigma_b^0$, which has not
been measured yet, is predicted to be
\ba%
m_{\Sigma_b^0} &\!\!\!=&\!\!\!
{1\over2}\left(m_{\Sigma_b^+}+m_{\Sigma_b^-} -\tilde{\beta}_4\right)
+ \Delta_{1b}^{\rm loop} \non\\
&\!\!\!=&\!\!\! 5810.3\pm1.8(\rm exp)\pm0.5(\rm th)~{\rm MeV}.
\ea%
We can also predict the mass difference between the $\Xi_b'^0$ and
$\Xi_b'^-$
\ba%
m_{\Xi_b'^0}-m_{\Xi_b'^-} &\!\!\!=&\!\!\!
{1\over2}\left(m_{\Sigma_b^+}-m_{\Sigma_b^-} -\tilde{\beta}_4\right) \non\\
&\!\!\!=&\!\!\! -4.0\pm1.8(\rm exp)\pm0.5(\rm th)~{\rm MeV}.
\ea%
In Table~\ref{tab:results}, the results of our work  are summarized
and a comparison with those obtained in quark models is given.

\section{Discussions and summary}

In this work, we have calculated the mass splittings within the
heavy baryon isospin multiplets $\Sigma_{c(b)}$ and $\Xi_{c(b)}'$ to
${\cal{O}}(p^3)$ in the chiral expansion. Our main results are given
in Eqs.~(\ref{eq:rescharm}--\ref{eq:resbottom}) and in
Table~\ref{tab:results}. To arrive at these results, we constructed
both the strong and the em Lagrangians at ${\cal{O}}(p^2)$ which are
responsible for the mass corrections. In contrast to mass splittings
in light quark baryon multiplets, there is an additional operator
that describes the hard virtual photons exchanged between the heavy
quark and light quarks accompanied by a LEC $\beta_{1h}$.
Remarkably, this term has a different sign for the charm baryons and
the bottom baryons. This is due to the fact that the sign of the
electric charge of the charm quark is different from that of the
bottom quark. It is the different interference between this term and
the other terms that drives the mass splittings within the
$\Sigma_c$ iso-triplet to have a different pattern compared to any
other known isospin multiplet. This leads one to expect that the
isospin mass splittings in the charm hadrons are always different
from those in the bottom hadrons even if the heavy quark symmetry
were exact. Besides the heavy baryons considered in this paper, the
$D$ and $B$--meson mass splittings,
$m_{D^\pm}-m_{D^0}=4.78\pm0.10$~MeV and
$m_{B^0}-m_{B^\pm}=0.37\pm0.24$~MeV~\cite{Amsler:2008zz} are a nice
example for the effect, although the ordering does not get changed
here.

There is no loop contribution to the mass splitting between the two
$\Xi_c'$ baryons, and we predict $m_{\Xi_c'^+} -
m_{\Xi_c'^0}=-0.2\pm0.6$~MeV. The present data for the masses of the
$\Xi_c'$ baryons are not accurate enough yet to test this prediction.
For the $\Sigma_b$ states, the $\beta_{1h}$ term interferes
constructively with the other terms and hence the loop corrections
are less important. The mass of the $\Sigma_b^0$ and the mass
difference $m_{\Xi_b'^0} - m_{\Xi_b'^-}$ are predicted to be
$5810.3\pm1.9~{\rm MeV}$ and $-4.0\pm1.9$~MeV, respectively, which
can be tested in future experiments.

\section*{Acknowledgments}

This work is partially supported by the Helmholtz Association
through funds provided to the virtual institute ``Spin and strong
QCD'' (VH-VI-231), by the EU Integrated Infrastructure Initiative
Hadron Physics Project under contract number RII3-CT-2004-506078
and by DFG (SFB/TR 16, ``Subnuclear Structure of Matter'').

\begin{appendix}

\section{Infrared regularization}
\label{app:ir}
\renewcommand{\theequation}{\thesection.\arabic{equation}}
\setcounter{equation}{0}

The finite masses of the baryons in the chiral limit spoil the
correspondence between the loop expansion and chiral expansion if
one uses  conventional dimensional
regularization~\cite{Gasser:1987rb}. Here we follow the infrared
regularization (IR) method developed by Becher and
Leutwyler~\cite{Becher:1999he} to overcome this problem. The IR
method has been extended to two loops~\cite{Schindler:2007dr} (see
also Ref.~\cite{Lehmann:2001xm}), and to the cases with spin-3/2
fields~\cite{Bernard:2003xf} and spin-1
fields~\cite{Bruns:2004tj,Bruns:2008ub} (for reviews, see
\cite{Scherer:2002tk,Bernard:2007zu}).

In the IR method, the scalar loop integral
\be%
H = {1\over i}\int \frac{d^dk}{(2\pi)^d} \frac{1}{(M^2-k^2)
\left[m^2-(P-k)^2\right]}.
\ee%
with $M$ and $m$ being the masses of the pion and baryon,
respectively, is split into two parts, one being infrared singular
and the other being infrared regular, $H=I+R$. Only the singular
part $I$, which is of order ${\cal{O}}(p)$, makes the expansion in
loops to coincide with the chiral expansion, hence leads to a
consistent power counting. The regular part $R$ can be expanded in
polynomials in $M$, hence it can be absorbed into the LECs order by
order.

In any regularization such as the IR which has a consistent power
counting for loops, the photon--baryon loops with each of the
vertices being of the ${\cal{O}}(e)$ order should be counted as
${\cal{O}}(e^2p) = {\cal O}(p^3)$. In the photon--baryon loops, when
taking $P^2=m^2$, which is necessary for calculating chiral
corrections to the baryon mass, there is no quantity of the order of
${\cal{O}}(p)$ since the photon is massless. Therefore, the
photon--baryon loops should vanish for calculating the mass shifts
of baryons. Such an argument is supported by explicit calculations
in the IR method.

The self-energy from any of the pion--baryon diagrams has the form
\be%
-i\Sigma^{\rm (n)}_{\Sigma_Q}(\not\!P;M,m) = - {g_i^2\over 4
F_\pi^2} \int {d^dk\over(2\pi)^d} \not\!k\gamma_5 {i\over
(k^2-M^2+i\varepsilon)}
{i(\not\!P-\not\!k+m)\over\left[(P-k)^2-m^2+i\varepsilon\right]}
\not\!k\gamma_5,
\ee%
where (n) is a diagram label, $P$ is the external momentum, and
$g_i~(i=1,2)$ are the $\Sigma_Q\Sigma_Q\pi$ and
$\Sigma_Q\Lambda_Q\pi$ ($Q=c,b$) coupling constants of the lowest
order Lagrangian, Eq.~(\ref{eq:L1}). After a few manipulations one
gets
\be%
\Sigma^{\rm (n)}_{\Sigma_Q}(\not\!P;M,m) = {g_i^2\over 4 F_\pi^2}
(\not\!P+m) \left[ M^2I(P^2) + (m-\not\!P)\not\!PI^{(1)}(P^2) -
\Delta_{\Sigma_Q} \right].
\ee%
In the IR method, the single baryon loop
$$\Delta_{\Sigma_Q}=i\int{d^dk\over(2\pi)^d}{1\over k^2-m^2}$$
vanishes. The expressions for the loop functions $I(P^2)$ and
$I^{(1)}(P^2)$ are given in Appendix~\ref{app:loop}.

Now let us focus on the $\Sigma_c^{++}$. Taking $\not\!P =
m_{\Sigma_c^{++}}$, the expressions for the diagrams
Fig.~\ref{fig:pionloop}(a) and (b) are very simple because the term
proportional to $(m-\not\!P)$ does not contribute,
\be%
\Sigma^{\rm (a)}(m_{\Sigma_c};M_0,m_{\Sigma_c}) = -
\frac{g_1^2M_{0}^3}{32\pi F_{\pi}^2},
\ee%
where the chiral limit mass $\cirm_{\Sigma_c}$ in the loop has been
replaced by the physical mass since the contribution of the
difference is of higher order. Note that up to ${\cal{O}}(p^3)$, we do not
need to distinguish the masses of baryons with different electric
charges in loops, so that $m_{\Sigma_c}$, instead of
$m_{\Sigma_c^{++}}$, is used for the arguments of the loop function.
Replacing the neutral pion mass $M_0$ by the charged pion mass
$M_{\pm}$, the expression for diagram (b) is obtained. The
contributions of the diagrams (a) and (b) do not depend on the
baryon mass.

The expression for diagram (c) is much more complicated since the
term proportional to $(m-\not\!P)$ has a finite contribution
\be%
\Sigma^{\rm (c)}(m_{\Sigma_c};M_{\pm},m_{\Lambda_c}) = {g_2^2\over 4
F_\pi^2} (m_{\Sigma_c}+m_{\Lambda_c}) \left[ M_{\pm}^2 {\bar
I}(m_{\Sigma_c}^2) + (m_{\Lambda_c}-m_{\Sigma_c})m_{\Sigma_c}{\bar
I}^{(1)}(m_{\Sigma_c}^2) \right],
\ee%
where the loop functions $I(P^2),I^{(1)}(P^2)$ are replaced by their
finite parts ${\bar I}(P^2),{\bar I}^{(1)}(P^2)$ (subtracting the
${\bar \lambda}$ parts), and the divergences can be absorbed in the
counterterms at  ${\cal{O}}(p^4)$  which are not considered here.
The expressions of the infrared singular parts of the loop integrals
are given in Appendix~\ref{app:loop}. Here we use the expansion of
the ${\bar I}(P^2)$ up to ${\cal{O}}(p)$. The chiral expansion of
$\bar I(P^2)$ up to ${\cal{O}}(p)$,
accounting for  the cut due to the opening of the $\Lambda_c\pi$ channel,
is
\ba%
{\bar I}(P^2) = - {\alpha\over 16\pi^2} \left\{\Omega(2\ln\alpha-1)
+ \sqrt{\Omega^2-1} \left[
\ln\left(\frac{\Omega+\sqrt{\Omega^2-1}}{\Omega-\sqrt{\Omega^2-1}}\right)
-2i\pi\right]\right\} + O(\alpha^2),
\ea%
where
\ba%
\alpha 
= {M_{\pm}\over m_{\Lambda_c}},\quad \Omega =
\frac{P^2-m_{\Lambda_c}^2-M_{\pm}^2}{2 M_{\pm} m_{\Lambda_c}}. \non
\ea%
For $P^2-m_{\Lambda_c}^2\sim {\cal O}(p)$, which is the case for
taking $P^2=m_{\Sigma_c}^2$ since
$m_{\Sigma_c}-m_{\Lambda_c}\simeq170$~MeV, $I^{(1)}(P^2)$ starts
from ${\cal{O}}(p^2)$, see Eq.~(\ref{appeq:I1}).
 The physical mass of the
$\Sigma_c$ is above the $\Lambda_c\pi$ threshold, correspondingly
$\Omega>1$.

Summing up the diagrams (a), (b) and (c), one gets the 
corrections to the mass of the $\Sigma_c^{++}$ at ${\cal{O}}(p^3)$
\be%
\Delta m_{\Sigma_c^{++}}^{\rm loop} =  -
\frac{g_1^2(M_0^3+M_{\pm}^3)}{32\pi F_{\pi}^2} + {\rm Re}\Sigma^{\rm
(c)}(m_{\Sigma_c};M_{\pm},m_{\Lambda_c}),
\ee%
where ${\rm Re}$ represents taking the real part. Similarly for the
$\Sigma_c^+$ and $\Sigma_c^0$, we have
\ba%
\Delta m_{\Sigma_c^{+}}^{\rm loop} &\!\!\!=&\!\!\! -
\frac{2g_1^2M_{\pm}^3}{32\pi F_{\pi}^2} + {\rm Re}\Sigma^{\rm
(c)}(m_{\Sigma_c};M_{0},m_{\Lambda_c}), \non\\
\Delta m_{\Sigma_c^{0}}^{\rm loop} &\!\!\!=&\!\!\! \Delta
m_{\Sigma_c^{++}}^{\rm loop}.
\ea%

\section{Loop integrals}
\label{app:loop}
\renewcommand{\theequation}{\thesection.\arabic{equation}}
\setcounter{equation}{0}

Defining
\ba%
\alpha &\!\!\!=&\!\!\! {M\over m},\quad s=P^2, \quad \Omega =
\frac{s-m^2-M^2}{2 M m},
\non\\
{\bar \lambda} &\!\!\!=&\!\!\! \frac{m^{d-4}}{(4\pi)^2}
\left[{1\over d-4} -
{1\over2}\left(\ln(4\pi)+\Gamma'(1)+1\right)\right], \non
\ea%
the infrared singular part of the loop integral
\be%
H(s) = {1\over i}\int \frac{d^dk}{(2\pi)^d} \frac{1}{(M^2-k^2)
\left[m^2-(P-k)^2\right]},
\ee%
is (the expression for $-1<\Omega<1$ is given in
Ref.~\cite{Becher:1999he})
\ba%
I(s) &\!\!\!=&\!\!\! {\bar I}(s) - \frac{s-m^2+M^2}{s}{\bar
\lambda}, \non\\
{\bar I}(s) &\!\!\!=&\!\!\! - {1\over 16\pi^2}
\frac{\alpha}{1+2\alpha\Omega+\alpha^2} \left[
(\Omega+\alpha)(2\ln\alpha-1) + F(s)\right],
\ea%
where
\ba%
F(s) = \begin{cases} \sqrt{\Omega^2-1} \left[
\ln\left(-\sqrt{\Omega^2-1}-\Omega-\alpha\right) -
\ln\left(\sqrt{\Omega^2-1}-\Omega-\alpha\right) \right], &
\Omega<-1,\\
2\sqrt{1-\Omega^2}
\arccos\left(-\frac{\Omega+\alpha}{\sqrt{1+2\alpha\Omega+\alpha^2}}\right),
& -1<\Omega<1,\\
\sqrt{\Omega^2-1} \left[
\ln\left(\sqrt{\Omega^2-1}+\Omega+\alpha\right) -
\ln\left(\Omega+\alpha-\sqrt{\Omega^2-1}\right) - 2i\pi \right], &
\Omega>1.
\end{cases}
\ea%
In a more compact way, one can rewrite this by keeping the $i\varepsilon$,
$\varepsilon \to 0^+$, explicitly,
which is necessary to choose the correct Riemann sheet,
\be%
F(s) = \sqrt{\Omega^2-1}
\left[\ln\left(-\sqrt{\Omega^2-1}-\Omega-\alpha-i\varepsilon\right)
-
\ln\left(\sqrt{\Omega^2-1}-\Omega-\alpha+i\varepsilon\right)\right].
\ee%

$I^{(1)}(s)$ is defined as
\be%
P^\mu I^{(1)}(s) = {1\over i}\int{d^dk\over(2\pi)^d} {k^\mu\over
(M^2-k^2) \left[m^2-(P-k)^2\right]}.
\ee%
One gets
\be%
\label{appeq:I1} I^{(1)}(s) = {1\over
2s}\left[(s-m^2+M^2)I(s)+\Delta_\pi-\Delta_{\Sigma_Q}\right],
\ee%
where in the IR method~\cite{Becher:1999he},
\ba%
\Delta_{\pi} &\!\!\!=&\!\!\! i\int{d^dk\over(2\pi)^d}{1\over
k^2-M^2} = 2M^2\left({\bar \lambda}+{1\over16\pi^2}\ln\alpha\right)~,
\non\\
\Delta_{\Sigma_Q} &\!\!\!=&\!\!\! i\int{d^dk\over(2\pi)^d}{1\over
k^2-m^2} = 0~.
\ea%

\end{appendix}

\end{document}